\documentclass[12pt]{article}
\usepackage{graphicx,amsmath}
\usepackage{graphicx,amsmath,lineno,color}

\usepackage{units}
\usepackage{caption} 
\usepackage{subcaption}
\usepackage[backend=biber,sorting=none,natbib=true,doi=true]{biblatex} %biblatex mit biber laden
\renewbibmacro{in:}{}
\usepackage[hidelinks,colorlinks=true,linkcolor=blue,urlcolor=blue,citecolor=blue]{hyperref}
\hypersetup{colorlinks=true}

\newlength{\dinwidth}
\newlength{\dinmargin}
\def\lapproxeq{\lower .7ex\hbox{$\;\stackrel{\textstyle                                                    
<}{\sim}\;$}}                                                    
\def\gapproxeq{\lower .7ex\hbox{$\;\stackrel{\textstyle                                                    
>}{\sim}\;$}}                                                    
\def\be{\begin{equation}}                                                    
\def\ee{\end{equation}}                                                    
\def\bea{\begin{eqnarray}}                                                    
\def\eea{\end{eqnarray}}

\def\sh{\hat s}
\def\sh2{{\hat s}^2}

\begin{document}                                                    
\titlepage                                                    
\begin{flushright}                                                    
%IPPP/19/??  \\                                                    
\today \\                                                    
\end{flushright} 
\vspace*{0.5cm}

\begin{center}                                                    
{\Large \bf About AKM scaling and oscillations in elastic scattering at very small momentum transfer at the LHC \\}

\vspace*{1cm}
% $\vec\nabla$                           
Per~Grafstr\"om  \\                       
\vspace*{0.5cm}                                                    
 Universit\`a di Bologna, Dipartimento di Fisica , 40126 Bologna, Italy\\
                        
\vspace*{1cm}                                                    
 
\begin{abstract}
The ATLAS and TOTEM collaborations have measured the differential elastic cross section at centre-of mass energy $\sqrt{s}$=13 TeV and at  small four-moment squared $|t|$. The data at very small $|t|$ i.e. $|t|<0.01GeV^{2}$ have been analysed in terms of so called AKM (Auberson, Kinoshita and Martin) oscillations. An indication of a possible oscillation of this type had previously been reported at $\sqrt{s}$=541 GeV using data from the UA4/2 experiment. There are no such indications in the data at 13 TeV examined here.   \end{abstract}

\end{center}

\vspace{1cm}

\section{Introduction}
Already  more than 50 years ago the possibility of  oscillations  in the forward direction of   the elastic differential cross section was discussed (see Ref.~\cite{PhysRevD.3.3185} and references therein).
Auberson, Kinoshita and Martin (AKM), assuming unequal particle-particle and particle-antiparticle cross section asymptotically,   proved within axiomatic field theory  that the scattering amplitude must have  infinitely many zeros in the forward direction. This fact does not in itself necessarily generates oscillations but certainly  opens up the possibility.     

Some indications of possible oscillations were reported in Ref.\cite{GAURON1997305}. The authors analyzed data from the UA4/2  experiment \cite{AUGIER1993448} at the SPS-collider at a centre-of-mass energy of $\sqrt s $=541 GeV. The UA4/2 experiment measured elastic antiproton-proton scattering in a $t$-range from -$t$=0.0007 $GeV^2$ to 0.12 $GeV^2$, where $t$ is the four-momentum squared.   The analysis reported in Ref.\cite{GAURON1997305} zoomed into the very small $t$-range ($-t<0.01$ $GeV^2$ )   of the UA4/2 data where AKM type oscillations might be present. A very simple model was used to identify a possible oscillatory behavior of the differential cross section. The model  introduces an extra term in the nuclear scattering amplitude $F_{N}$  which allows for possible oscillation that are compatible with the AKM theorem \cite{PhysRevD.3.3185}.
The nuclear amplitude then becomes
\begin{equation}
\label{nuclamp}
F_{N}=F_{Nexpo}+F_{Nosc}.
\end{equation}
Here $F_{Nexpo}$ is the standard nuclear scattering amplitude with an exponential $t$-dependence used by UA4/2 in their fits
\begin{equation}
\label{nucexp}
F_{Nexpo}=(\rho+i)\sigma_{T}e^{bt/2} .
\end{equation}
The $\rho $-parameter is defined as $\rho=\frac{ReF_{el}}{ImF_{el}}$ extrapolated to $t$=0, $\sigma_{T}$ is the total cross section and $b$ is the so called slope parameter.
The extra oscillatory term was taken as
\begin{equation}
\label{nuclamposc}
F_{Nosc}= (A+Bi) \cdot (sin\tau)/\tau.
\end{equation}
Here $\tau$ is the scaling variable introduced by AKM
\cite{PhysRevD.3.3185}. 
\begin{equation}
\label{tau}
\tau= \sqrt{-t/t_{0}}\cdot ln(s/s_{0}),
\end{equation}
and where $s_{0}$ is set to 1 $GeV^{2}$.
This scaling implies that the $t$-behaviour and the $s$-behaviour are linked when the centre-of-mass energy growths.
The parameters A,B and $t_{0}$ are obtained by a fit to the UA4/2 data.

The main result of the analysis in Ref.\cite{GAURON1997305} is the fact that $\chi^{2}/dof$ decreases with 10~\% when the extra oscillatory term is added. Strong indications of the presence of oscillations are also confirmed by a statistical analysis. The authors conclude  that there are possible evidence for AKM structure at low~$t$ in the UA4/2 data and  look forward to see if this can be confirmed when data at higher $\sqrt s$ is available.  Today there exist precise data both from ATLAS \cite{ATLAS:2022mgx}  and TOTEM\cite{TOTEM:2017sdy} at very small $t$-values and at $\sqrt s$=13~TeV. It is the purpose of this note to see if the possible AKM structure seen at $\sqrt s$=541~GeV also is present at the LHC with $\sqrt s$=13~TeV.
\section{ Re-analysis of the UA4/2 data}
\begin{figure}[ht]
\begin{subfigure}{0.49\textwidth}
\includegraphics[width=0.9\linewidth, height=6cm]{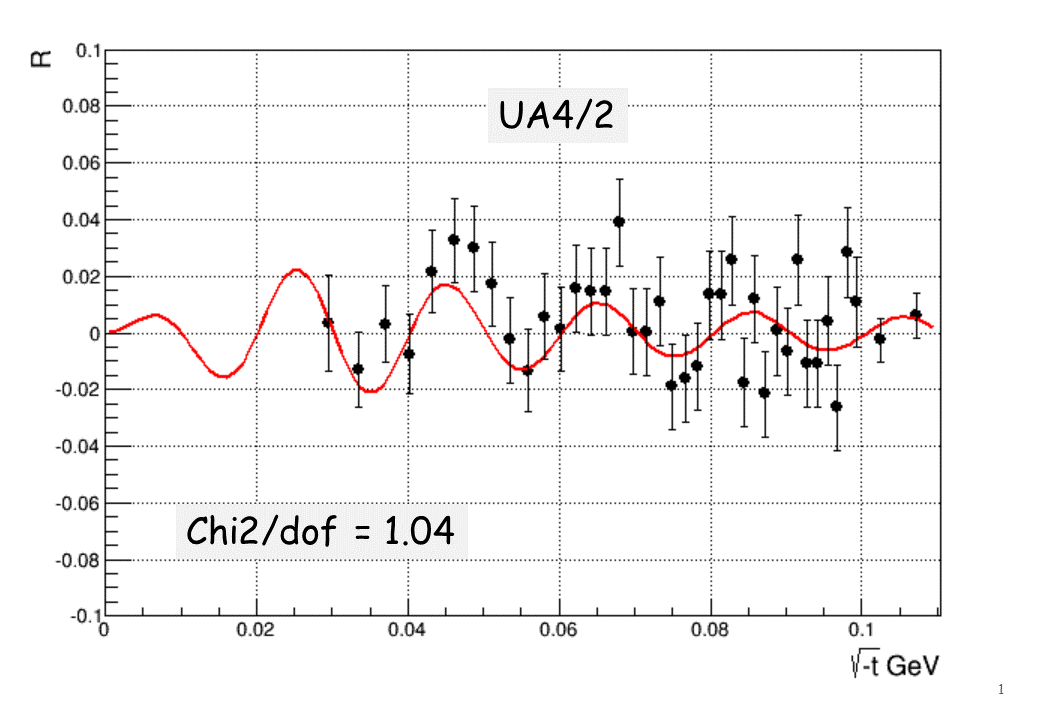} 
\caption{}
\label{fo1}
\end{subfigure}
\begin{subfigure}{0.49\textwidth}
\includegraphics[width=0.9\linewidth, height=6cm]{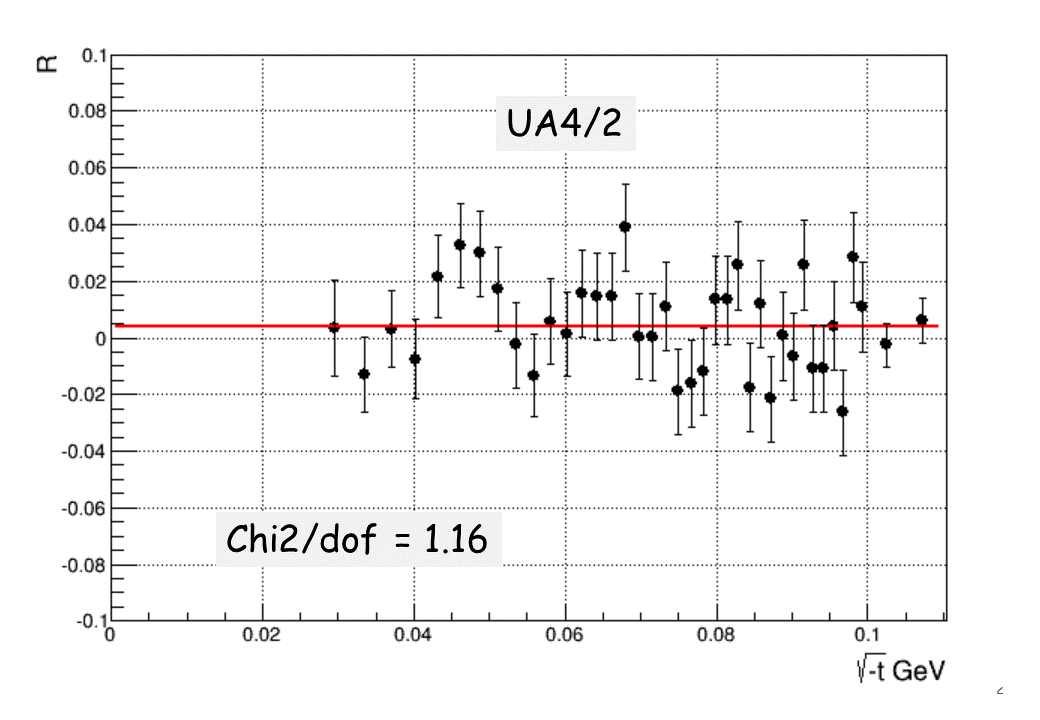}
\caption{ }
\label{fo2}
\end{subfigure}
    \caption{ The ratio R  (defined in the text and in  Eq.~\ref{defR}) as a function of $\sqrt{(-t)}$ . The data points are taken from the UA4/2 data. The  curve in (a) corresponds to the oscillatory model and in (b) to the case of no oscillations.}
\label{ua4}
\end{figure}

To be sure that the oscillatory model  given by Eqs.~\ref{nuclamp}-\ref{tau} is correctly employed for the ATLAS and TOTEM data, the UA4/2 data have, as a starting point, been  reanalyzed  and  compared with the result given in \cite{GAURON1997305}. Like in~\cite{GAURON1997305} the ratio R is defined as :
\begin{equation}
\label{defR}
R=\frac{\frac{d\sigma_{el}}{dt}(data)}{\frac{d\sigma_{el}}{dt}(expo)}-1.
\end{equation}
where \(\frac{d\sigma_{el}}{dt}(data)  \) is the elastic differential cross section as measured by UA4/2   and \(\frac{d\sigma_{el}}{dt}(expo) \) represents the parametrization of the cross section used by the UA4/2 collaboration employing the nuclear amplitude given by Eq.~\ref{nucexp}. The result is summarized  in Figure~\ref{ua4}. The data points in the figure  corresponds to R calculated according to Eq.~\ref{defR}. The  curve in Figure~\ref{fo1} represents the calculation of R using the oscillatory model. In Figure~\ref{fo2} is shown the result if no oscillations are introduced. The result shown in Figure~\ref{fo1} is completely identical to the result presented in Figure 2 in Ref~\cite{GAURON1997305}. Moreover the increase of $\chi^{2}/dof$ assuming no oscillation agrees with the 10\% reported in~\cite{GAURON1997305}. Thus we conclude that  the oscillatory model is employed in a correct way here.

\section {Analysis of the recent ATLAS and TOTEM data at \texorpdfstring{$\sqrt s$}{Lg}=13 TeV. }

\begin{figure}[htb]
\begin{subfigure}{0.49\textwidth}
\includegraphics[width=0.9\linewidth, height=6cm]{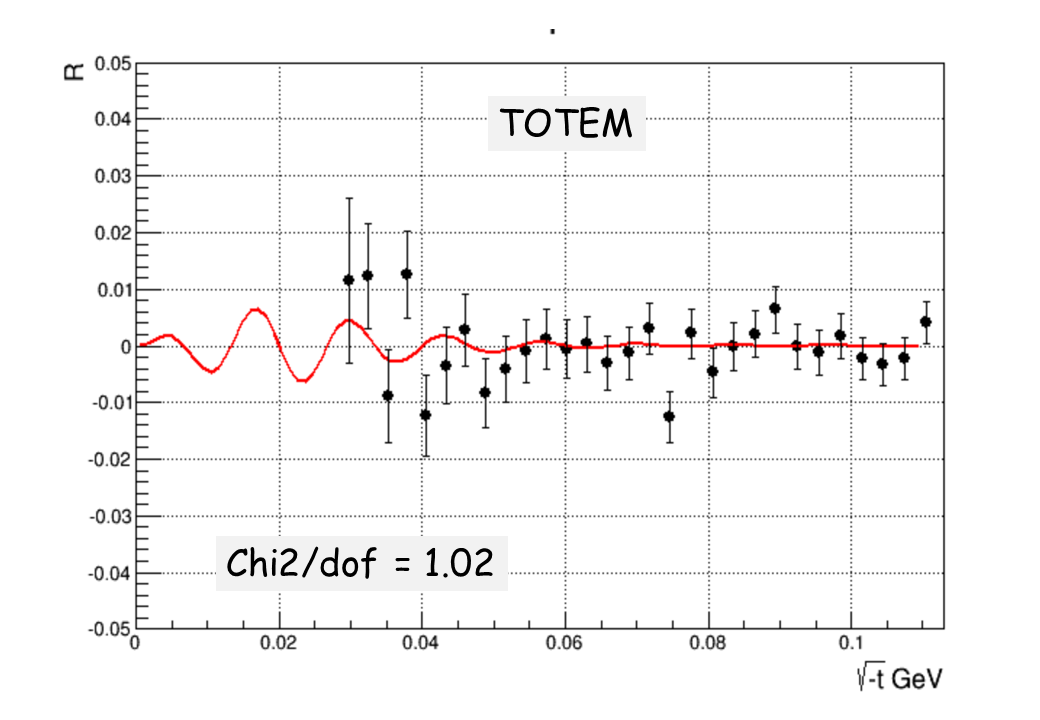} 
\caption{ }
\label{fo3}
\end{subfigure}
\begin{subfigure}{0.49\textwidth}
\includegraphics[width=0.9\linewidth, height=6cm]{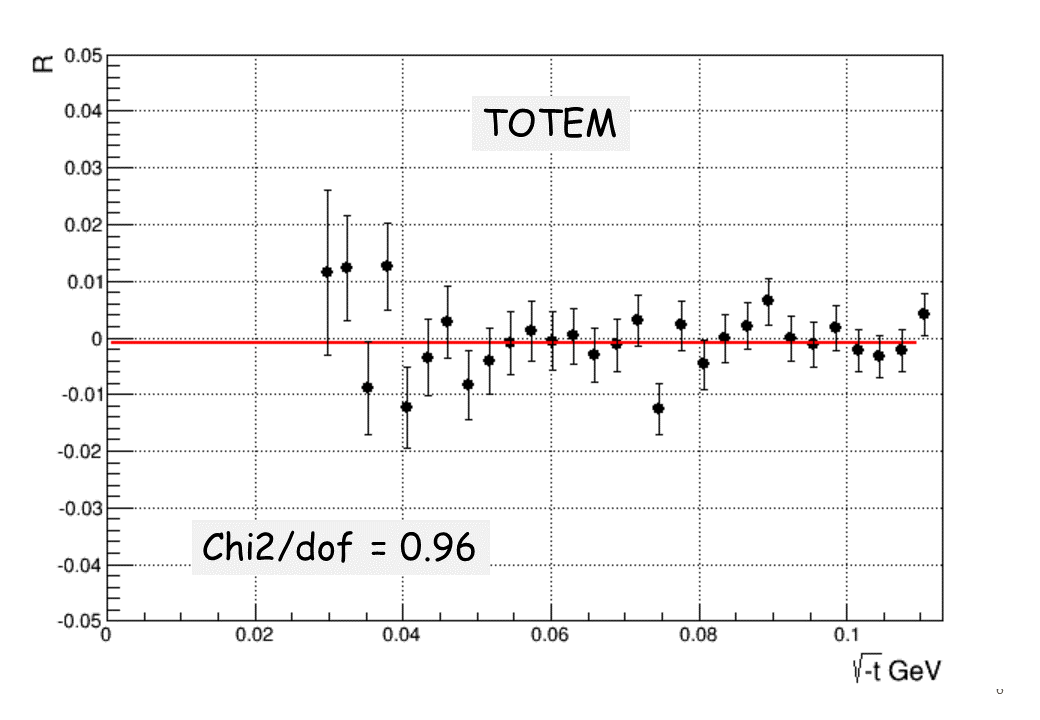}
\caption{ }
\label{fo4}
\end{subfigure}
\caption{The ratio R  (defined in the text and in  Eq.~\ref{defR}) as a function of $\sqrt{(-t)}$ . The data points are taken from the TOTEM data. The curve in (a) corresponds to the oscillatory model and in (b) to the case of no oscillations. }
\label{totem}
\end{figure}

\begin{figure}[b]
\begin{subfigure}{0.49\textwidth}
\includegraphics[width=0.9\linewidth, height=6cm]{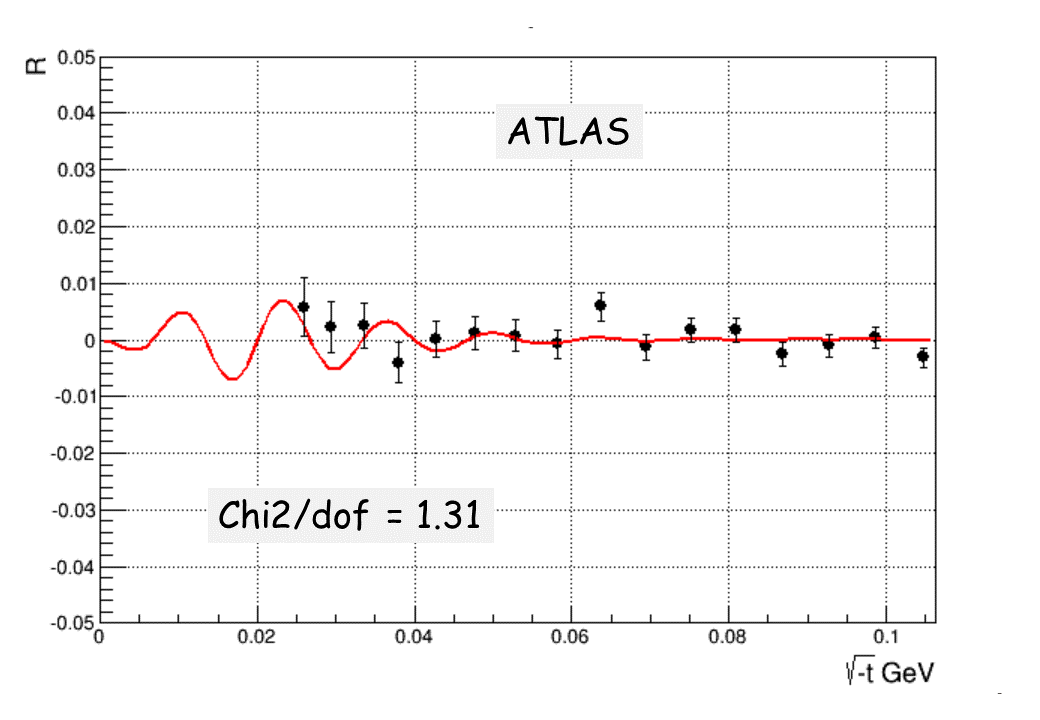} 
\caption{ }
\label{fo5}
\end{subfigure}
\begin{subfigure}{0.49\textwidth}
\includegraphics[width=0.9\linewidth, height=6cm]{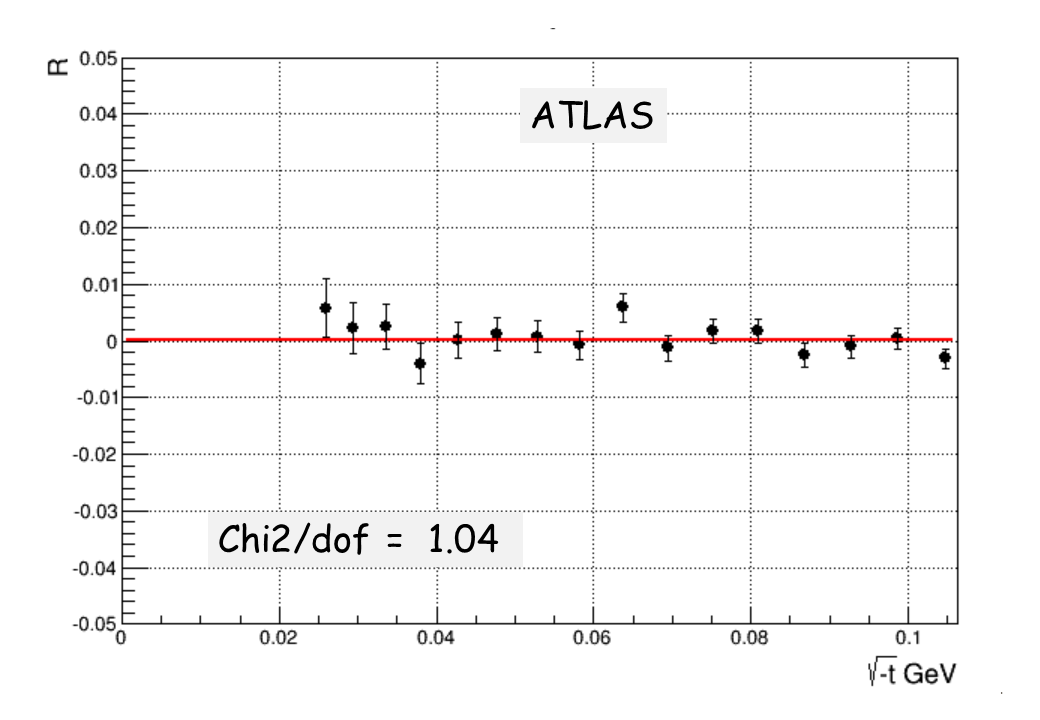}
\caption{ }
\label{fo6}
\end{subfigure}
\caption{The ratio R  (defined in the text and in  Eq.~\ref{defR}) as a function of $\sqrt{(-t)}$ . The data points are taken from the ATLAS data. The  curve in (a) corresponds to the oscillatory model and in (b) to the case of no oscillations.}
\label{atlas}
\end{figure}

The recent ATLAS \cite{ATLAS:2022mgx}  and TOTEM\cite{TOTEM:2017sdy} data have been analyzed  in the same manner as the UA4/2 data. The parameters entering into the standard nuclear amplitude $F_{Nexpo}$ (Eq.\ref{nucexp}) are taken from fits performed by respective experiments. For the value of $t_{0}$  the value found in the fit to the UA4/2 data i.e. $t_{0}$= 0.0016 $GeV^{2}$ is used. Thus it is  assumed that the AKM scaling of Eq.\ref{tau} is valid going from $\sqrt s$=540 GeV to 13 TeV. The parameter's A and B in $F_{Nosc}$ defined by Eq~\ref{nuclamposc}. are left free in the fit. The results are shown in Figure~\ref{totem} for TOTEM and Figure~\ref{atlas} for ATLAS. In the case of TOTEM there is no significant difference in $\chi^{2}$/dof between the two hypothesis. In the case of ATLAS the no oscillation hypothesis gives a slightly  lower $\chi^{2}$/dof.

\clearpage
\section {Consequences for the \texorpdfstring{$\rho$}{Lg}-parameter.}
\begin{figure}{}
\centering
\includegraphics[width=0.6\textwidth]{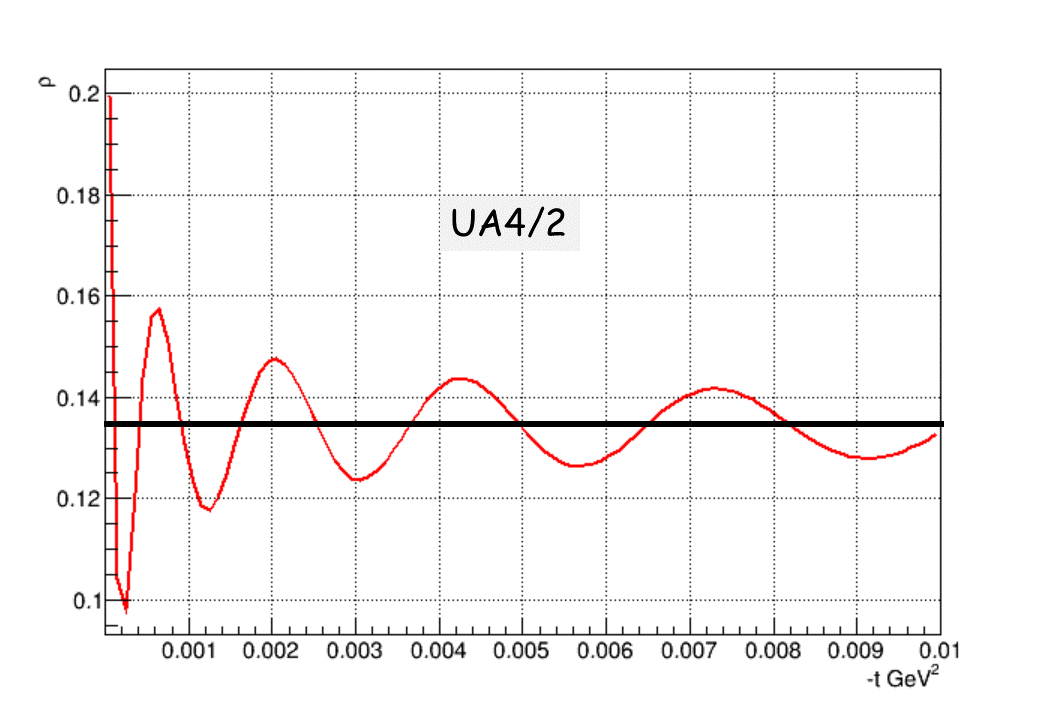}
\caption{\label{fo7} The $\rho$-parameter as a function of $t$ assuming an oscillatory term as part of the nuclear amplitude. Here is plotted the case for UA4/2 at $\sqrt{s}$=541 GeV}.
\end{figure}
\begin{figure}[ht]
\begin{subfigure}{0.49\textwidth}
\includegraphics[width=0.9\linewidth, height=6cm]{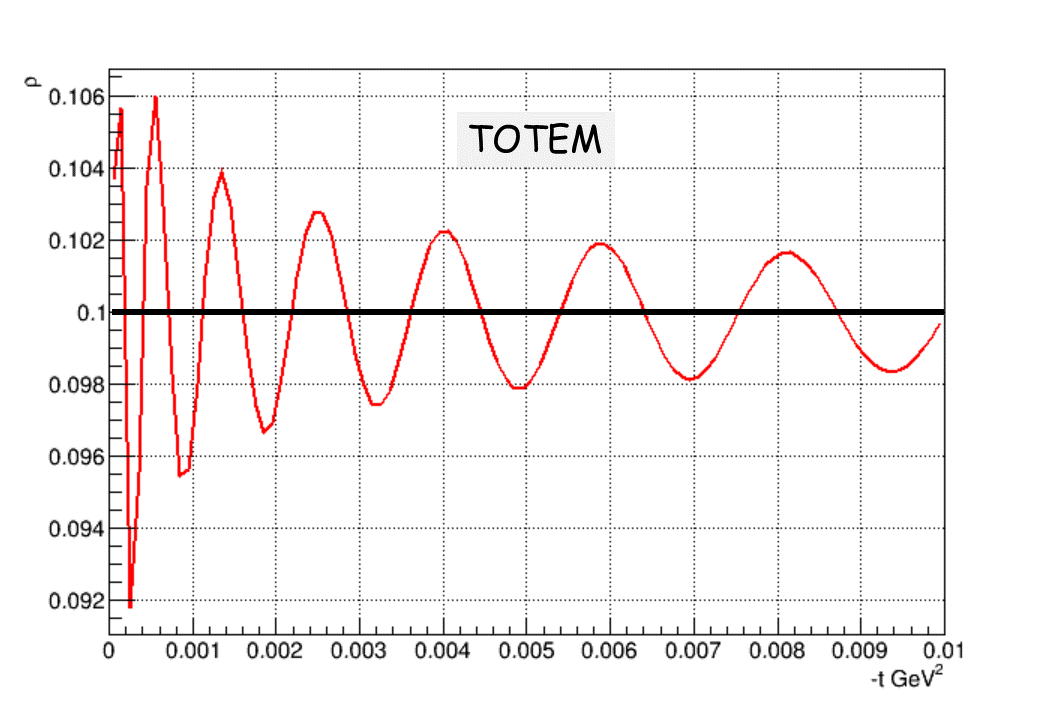} 
\caption{ $TOTEM$}
\label{fp3}
\end{subfigure}
\begin{subfigure}{0.49\textwidth}
\includegraphics[width=0.9\linewidth, height=6cm]{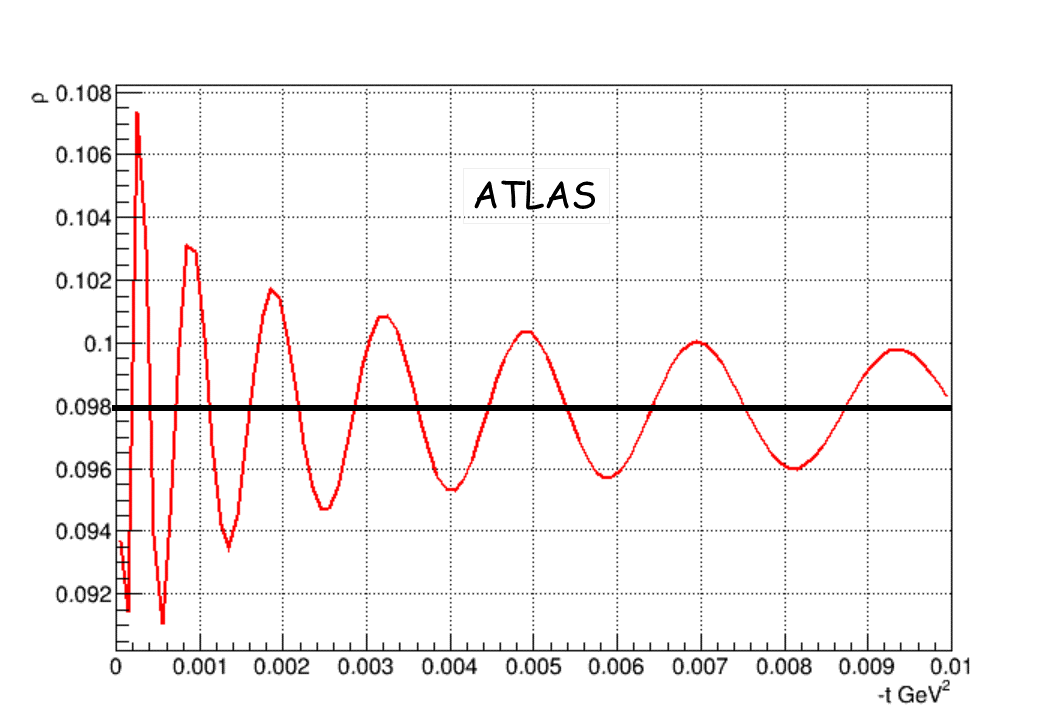}
\caption{$ATLAS$ }
\label{fp4}
\end{subfigure}
\caption{The $\rho$-parameter as a function of $t$ assuming an oscillatory term as part of the nuclear amplitude. Here is plotted the case for TOTEM and ATLAS at $\sqrt{s}$=13 TeV.}
\label{attoro}
\end{figure}
It is clear that if an oscillatory term of the type proposed in Eq.~\ref{nuclamposc} would really exist, it would have as a consequence  that $\rho (t)=\frac{ReF_{el}(t)}{ImF_{el}(t)}$ would vary in an oscillatory manner as a function of $t$. This was investigated in Ref.~\cite{GAURON1997305} and is also investigated here. The result for the UA4/2 data is shown in Figure~\ref{fo7}. It is pointed out in Ref.~\cite{GAURON1997305} that if the size of the oscillatory term found in the fit to the UA4/2 data would correspond to reality then the extrapolation of $\rho$ to $t$=0 would be problematic. This is also clearly seen in the result here  shown in Figure~\ref{fo7}. Observe that the nominal value  of the $\rho$-parameter at $t$=0 found by the UA4/2 collaboration is $\rho$=0.135 and that $\rho(t)$ oscillates around this value.

 The same analysis has been performed  on the ATLAS and TOTEM data. The results are shown in Figure~\ref{attoro}. As the oscillatory term here are smaller the extrapolation problem is not really an issue.
 The nominal values found  by ATLAS and TOTEM for  $\rho ~(t=0) $ are 0.098 and 0.10 respectively and the curves oscillate around those values. Of course those oscillations are irrelevant as there is no  experimental indication  of an oscillatory term being present at 13 TeV.

\section{Conclusion}
 The indications of possible oscillations in the differential elastic cross  sections as measured by the UA4/2 collaboration at $\sqrt{s}$=541 GeV and very small values of $-t$ i.e below $0.01~GeV^{2}$ was  taken as a starting point. Assuming that AKM scaling holds the data from the ATLAS and the TOTEM collaboration at $\sqrt{s}$=13 TeV have been analyzed.  There 
 are no indications of oscillations in those data. 
 
 It should be pointed out that we do not discuss the possibility of oscillations at higher $t$ values and with much longer wavelength. The possible existence  of such types of oscillations at the LHC, which are not of AKM nature, has been discussed by Oleg Selyugin in a number of papers, see Refs~
\cite{Selyugin_2019},\cite{Selyugin_2021},\cite{selyugin2022anomalies} and \cite{selyugin2023new}.

\vspace{\baselineskip} 
{\bf\Large Acknowledgements}

Many thanks to Fares Djama and Maurice Haguenauer that made us aware of previous discussions of oscillations related to the UA4/2 data. Also many thanks to  Valery Khoze, Misha Ryskin and  Oleg Selyugin for reading and  commenting this note. In addition, thanks to the members of ATLAS-ALFA group with whom I have discussed this topic many times.
\vspace{\baselineskip} 

%\bibliographystyle{unsrt}
%\bibliography{AKM.bib}
\printbibliography 

@article{PhysRevD.3.3185,
  title = {Violation of the Pomeranchuk Theorem and Zeros of the Scattering Amplitudes},
  author = {Auberson, G. and Kinoshita, T. and Martin, A.},
  journal = {Phys. Rev. D},
  volume = {3},
  issue = {12},
  pages = {3185--3194},
  numpages = {0},
  year = {1971},
%  month = {Jun},
  publisher = {American Physical Society},
  doi = {10.1103/PhysRevD.3.3185},
  url = {https://link.aps.org/doi/10.1103/PhysRevD.3.3185}
}

@article{GAURON1997305,
title = {The AKM theorem and oscillations in the hadron scattering amplitude at high energy and small momentum transfer},
journal = {Physics Letters B},
volume = {397},
number = {3},
pages = {305-310},
year = {1997},
issn = {0370-2693},
doi = {https://doi.org/10.1016/S0370-2693(97)00171-8},
url = {https://www.sciencedirect.com/science/article/pii/S0370269397001718},
author = {P. Gauron and B. Nicolescu and O.V. Selyugin},

}

@article{AUGIER1993448,
title = {A precise measurement of the real part of the elastic scattering amplitude at the SppS},
journal = {Physics Letters B},
volume = {316},
number = {2},
pages = {448-454},
year = {1993},
issn = {0370-2693},
doi = {https://doi.org/10.1016/0370-2693(93)90350-Q},
url = {https://www.sciencedirect.com/science/article/pii/037026939390350Q},
author = {C. Augier and D. Bernard and J. Bourotte and M. Bozzo and A. Bueno and R. Cases and F. Djama and M. Haguenauer and V. Kundrát and M. Lokajíček and G. Matthiae and A. Morelli and F. Natali and S. Němeček and M. Novák and E. Sanchis and G. Sette and M. Smižanská and J. Velasco}
}

@article{ATLAS:2022mgx,
    author = "Aad, Georges and others",
    collaboration = "ATLAS",
    title = "{Measurement of the total cross section and $\rho $-parameter from elastic scattering in pp collisions at $\sqrt{s}=13$~TeV with the ATLAS detector}",
    eprint = "2207.12246",
    archivePrefix = "arXiv",
    primaryClass = "hep-ex",
    reportNumber = "CERN-EP-2022-129",
    doi = "10.1140/epjc/s10052-023-11436-8",
    journal = "Eur. Phys. J. C",
    volume = "83",
    number = "5",
    pages = "441",
    year = "2023"
}

@article{TOTEM:2017sdy,
    author = {{TOTEM Collaboration}},
    title = "{First determination of the ${\rho }$ parameter at ${\sqrt{s} = 13}$ TeV: probing the existence of a colourless C-odd three-gluon compound state}",
    eprint = "1812.04732",
    archivePrefix = "arXiv",
    primaryClass = "hep-ex",
    reportNumber = "CERN-EP-2017-335, CERN-EP-2017-335-v3",
    doi = "10.1140/epjc/s10052-019-7223-4",
    journal = "Eur. Phys. J. C",
    volume = "79",
    number = "9",
    pages = "785",
    year = "2019"
}

@article{Selyugin_2019,
	doi = {10.1016/j.physletb.2019.134870},
  
	url = {https://doi.org/10.1016%2Fj.physletb.2019.134870},
  
	year = 2019,
%	month = {oct},
  
	publisher = {Elsevier {BV}
},
  
	volume = {797},
  
	pages = {134870},
  
	author = {O.V. Selyugin},
  
	title = {New feature in the differential cross sections at 13 {TeV} measured at the {LHC}},
  
	journal = {Physics Letters B}
}

@article{Selyugin_2021,
	doi = {10.1142/s0217732321501480},
  
	url = {https://doi.org/10.1142%2Fs0217732321501480},
  
	year = 2021,
%	month = {jun},
  
	publisher = {World Scientific Pub Co Pte Ltd},
  
	volume = {36},
  
	number = {18},
  
	pages = {2150148},
  
	author = {O. V. Selyugin},
  
	title = {Anomaly in the differential cross sections at 13 {TeV}
},
  
	journal = {Modern Physics Letters A}
}

@misc{selyugin2022anomalies,
      title={Anomalies in the differential cross sections at 13 TeV}, 
      author={O. V. Selyugin},
      year={2022},
      eprint={2201.02403},
      archivePrefix={arXiv},
      primaryClass={hep-ph}
}

@misc{selyugin2023new,
      title={New features in the differential cross sections measured at the LHC}, 
      author={O. V. Selyugin},
      year={2023},
      eprint={2210.02970},
      archivePrefix={arXiv},
      primaryClass={hep-ph}
}

\end{document}